\documentclass[aip,pof,reprint,onecolumn]{revtex4-1}
\usepackage{amsmath}
\usepackage{graphicx}
\usepackage{color}
\usepackage{hyperref}
\usepackage{stmaryrd}
\usepackage{amssymb}
\usepackage{wasysym}
\usepackage[utf8]{inputenc}
\usepackage{color}

\begin{document}

\title{Cross-waves induced by the vertical oscillation of a fully immersed vertical plate}

\author{Fr\'{e}d\'{e}ric Moisy}
\author{Guy-Jean Michon}
\altaffiliation{Present address: Institut Jean Le Rond d'Alembert, 4 place Jussieu, 75252 Paris, France.}
\author{Marc Rabaud}
\author{Eric Sultan}
\affiliation{Universit\'e Paris-Sud, UPMC Universit\'e Paris 6, CNRS, Laboratoire FAST. B\^atiment 502, 91405 Orsay, France.}

\date{\today}

\begin{abstract}

Capillary waves excited by the vertical oscillations of a thin elongated plate below an air-water interface are analyzed using time-resolved measurements of the surface topography. A parametric instability is observed above a well defined acceleration threshold, resulting in a so-called cross-wave, a staggered wave pattern localized near the wavemaker and oscillating at half the forcing frequency. This cross-wave, which is stationary along the wavemaker but propagative away from it, is described as the superposition of two almost anti-parallel propagating parametric waves making a small angle of the order of $20^\mathrm{o}$ with the wavemaker edge. This contrasts with the classical Faraday parametric waves, which are exactly stationary because of the homogeneity of the forcing. Our observations suggest that the selection of the cross-wave angle results from a resonant mechanism between the two parametric waves and a characteristic length of the surface deformation above the wavemaker.

\end{abstract}

\maketitle

\section{Introduction}

When surface waves are excited by a wavemaker, 
localized transverse stationary waves, with crests approximately normal to the wavemaker, are frequently encountered
in addition to the expected longitudinal propagative waves.
These transverse waves, which received the name of {\it cross-waves}, oscillate at half the forcing frequency, indicating a parametric instability mechanism. They have been first described in the pioneering work of Faraday,\cite{Faraday1831} who generated them with a vertical plate or cylindrical cork oscillating at large frequency at the surface of water. Capillary cross-waves have been observed with a variety of excitation sources: vertically oscillating horizontal cylinder, sphere  or wedge, or radially oscillating vertical cylinder.\cite{Schuler1933,Taneda1986,Taneda1991,Taneda1994,Taneda1995,Becker1991,Krasnopolskaya1996} This phenomenon is also commonly observed for gravity waves generated by paddle-type wavemakers oscillating at low frequency in large tanks.\cite{Barnard1972,Lichter1986,Shemer1987,Shemer1990}

For both capillary and gravity waves, low amplitude forcing first generates longitudinal waves at the forcing frequency and propagating away from the wavemaker. As the forcing amplitude is increased above a certain threshold, cross-waves appear as a modulation of the primary longitudinal waves,  but they remain localized close to the wavemaker, and their amplitude may become much larger than the amplitude of the primary longitudinal wave.\cite{Garrett1970} This indicates that the 
cross-waves are not a secondary instability of the primary waves, 
but their energy is rather directly pumped from the flow induced by the motion of the wavemaker.\cite{Krasnopolskaya1996, Barnard1972, Mahony1972} Cross-waves may be therefore considered as a particular instance of the general problem of trapped waves,\cite{Longuet1967,McIver1985} in which an immersed disturbance or a change in topography act as a waveguide. In the case of trapped waves along a shore line, also called edge waves,\cite{Johnson2007} their energy is supplied by the incoming waves from the open sea. Inversely, in the cross-wave problem, energy is supplied by the wavemaker, and remains trapped in its close vicinity.

For inviscid gravity waves in a tank of short length, Garrett\cite{Garrett1970} showed that a Mathieu's equation could be
derived for the cross-wave amplitude. This equation classically describes a parametric resonance in a forced oscillator, and results in the growth of a subharmonic response. Since the damping length is large for gravity waves, the excited primary waves are usually standing waves resonant with the length of the tank, whereas the cross-waves are exactly normal to the primary wave, with a wavelength discretized by the width of the tank. The detuning between the excitation frequency and  the resonant frequencies of the tank is therefore an important parameter in this problem. This approach has been extended for an infinite tank by Mahony\cite{Mahony1972} and Jones,\cite{Jones1984} who stressed the role of the non-propagating oscillating flow near the wavemaker in the selection of the resonant parametric wave. The stability of the cross-waves has been further investigated in the case of plane waves\cite{Miles1988} and circular waves.\cite{Becker1991}

Miles\cite{Miles1990} reviewed several types of parametrically forced surface waves, and examined the connection between cross-waves and the Faraday instability. In the classical Faraday experiment, a thin layer of a liquid is vertically oscillated, inducing a homogeneous forcing described by a modulation of the gravity.\cite{Benjamin1954} This experiment has received considerable interest in the last two decades, as it provides a simple system which exhibits nonlinear pattern formation and transition to spatiotemporal chaos. Although the localized cross-wave forcing investigated in the present paper differs from the homogeneous forcing in the Faraday experiment, these two parametric instabilities share a number of similar properties. In particular, in both cases viscosity is responsible for the finite threshold in forcing amplitude.\cite{Cerda1998,Bernoff1989} The essential difference is that the homogeneity of the forcing in the Faraday configuration induces a strictly stationary pattern, originating from a set of excited parametric waves of equal amplitude and wavevectors summing up to zero, whereas a partially stationary pattern (i.e., stationary along the wavemaker and propagating normal to it) is allowed in general for a localized forcing.

In this paper we characterize in detail the transition and the spatiotemporal properties of the capillary cross-waves excited by a fully immersed oscillating plate in a large water tank. The originality of the present work is that the wavemaker has no contact line with the water surface, resulting in negligible generation of longitudinal primary waves.
The resulting nearly pure cross-wave pattern is analyzed from time-resolved measurements of the free surface topography, obtained by the Free-Surface Synthetic Schlieren method (FS-SS).\cite{Moisy2009} This optical method provides the instantaneous two-dimensional surface height with a vertical resolution of 2~$\mu$m, a major improvement compared to one-point time series of surface depth given by conventional gauge measurements. We show that the cross-waves in this system are not exactly stationary, and not exactly normal to the wavemaker, but they rather systematically form a well defined angle with the wavemaker. The origin of this angle is discussed in the frame of a resonance with the non-propagating oscillating flow localized near the wavemaker.

\section{Experimental setup}
\label{sec:exp}

\subsection{The immersed wavemaker}

\begin{figure}
\centerline{\includegraphics[width=8cm]{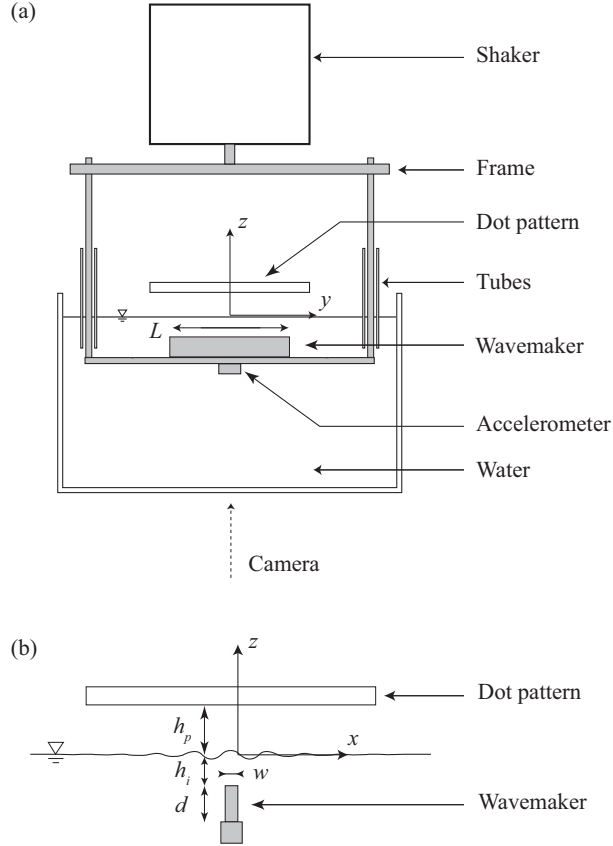}}
\caption{Experimental setup. (a) Front view. The wavemaker is hold on a rigid frame, and oscillated vertically below the free surface. (b) Side view of the wavemaker, with a sketch of the longitudinal primary waves, propagating in $x$ directions on both sides of the wavemaker; FS-SS measurements are performed only for $x>0$.\label{fig:setup}}
\end{figure}

The experimental setup, sketched in figure \ref{fig:setup}, consists in a glass tank, of size 60 cm $\times$ 30 cm, filled with 20~cm of water. The wavemaker is a thin plastic rectangular plate, of length $L$, height $d=20$~mm and thickness $w=5$~mm, located at a variable depth $h_i$ below the free surface. Two lengths, $L=120$ and 240~mm, have been used, and the immersion depth $h_i$ has been varied between 2 and 10~mm. The wavemaker is along the $y$ direction, with $-L/2 \leq y \leq L/2$. Longitudinal primary waves are propagating along the $x$ direction on both sides of the wavemaker. The tank is significantly larger than the viscous damping length of the waves, so the waves can be considered as effectively unbounded. Measurements are restricted here on one side only, $x>0$, with $x=0$ taken at the edge of the wavemaker.

The wavemaker is fixed to a stainless steal frame, and vertically oscillated by an electromagnetic shaker (Sinocera JZK-10) located above the water tank. 
In order to avoid residual generation of waves where the two lateral rods of the frame cross the water surface, the lateral rods are enclosed in fixed vertical tubes.
The top of the wavemaker is oscillating as $z(t) = -h_i + a_0 \cos \omega t$, with $z=0$ taken at the free surface. The oscillation frequency $f = \omega / 2\pi$ has been varied between 20 and 60~Hz, and the maximum amplitude is $a_0 = 2$~mm. The non-dimensional acceleration,
$$
\Gamma = a_0 \omega^2 / g,
$$
where $g$ is gravity, lies in the range 0 to 4. Note here that using $g$ to normalize the acceleration is only a convenient choice, the onset acceleration for cross-waves being {\it a priori} not directly related to $g$. The wavemaker acceleration is measured by an accelerometer (Sensel Measurement BDK10) fixed just below the wavemaker, with a sensitivity of 43.7 mV/$g$.  We have checked that no significant harmonics are present in the accelerometer signal for the explored range of frequency. 

The tank is filled with tap water at room temperature, and the measurements are performed during the two hours following the filling. This procedure ensures a minimal pollution of the free surface, and thus a limited evolution of the air-water surface tension. The actual surface tension is determined from the measurement of the wavelength of the longitudinal primary wave (see Sec.~\ref{sec:Longitudinal}), and is found to be $\gamma = (60 \pm 2) \times 10^{-3}$~N~m$^{-1}$. The resulting capillary-gravity crossover frequency is $f_c = (4 \rho g^3  / \gamma)^{1/4} / (2\pi) \simeq 14.2$~Hz (with $\rho$ the water density), so that the waves generated by the wavemaker in the range $f = 20 - 60$~Hz are essentially capillary waves, although gravity effects cannot be neglected.

\subsection{Surface height measurements}
\label{sec:Surf_Measure}

Two types of measurements have been performed:  One-point slope measurement, using a reflected laser beam, and whole-field instantaneous topography measurements from Free-Surface Synthetic Schlieren (FS-SS).\cite{Moisy2009}

The one-point slope measurement is useful to determine the growth rate and the onset curve of the cross-wave instability. A laser beam, of diameter 1~mm, is reflected on the surface near the wavemaker with an incidence angle of $5^\mathrm{o}$, and the reflected beam intersects a screen located at a distance of 4~m from the measurement point. A slope resolution of $10^{-4}$~rad can be achieved, yielding a typical amplitude resolution of 0.2~$\mu$m for the wavelengths considered here. 

The principle of FS-SS for the measurement of the surface topography is detailed in Moisy {\it et al.},\cite{Moisy2009} and is only briefly described here. This method is based on the analysis of the refracted image of a reference pattern visualized through the deformed interface. This approach, and similar ones developed on the same basis,\cite{Kurata1990,Elwell2004,Fouras2008} 
can be seen as a particular case of Synthetic Schlieren\cite{Dalziel2000} for a stepwise variation of the refraction index at the surface, allowing for a complete reconstruction of the surface height field.

A dense pattern of randomly distributed dots, printed on a $10 \times 15$~cm$^2$ overhead transparency, is located at a distance $h_p$ above the free surface (see figure~\ref{fig:setup}b). This pattern is illuminated using an homogeneous lighting plate, and is imaged by a camera located below the water tank, through the free surface and the transparent bottom of the tank. Images of the pattern are acquired through the surface deformed by the waves, and are compared to the reference image obtained for a flat surface. The apparent displacement field $\delta {\bf r}(x,y)$ between the reference image and distorted images induced by the light refraction through the deformed interface is determined using a digital cross-correlation algorithm.\cite{Davis,pivmat} In the paraxial approximation (camera located far from the interface), and in the limit of small slopes and small amplitudes, the displacement field is simply proportional to the gradient of the surface height, $\delta {\bf r}(x,y) = - (1-1/n) h_p \nabla \eta$, where $h_p$ is surface-pattern distance and $n$ is the water refraction index. The surface deformation $\eta(x,y)$ is then reconstructed from integration of $\delta {\bf r}(x,y)$, using a least-square inversion of the gradient operator.\cite{pivmat}

The images are acquired by a high-speed camera, $1280 \times 1024$~pixels, operating at 500 frames per second. Sequences of 1~s are recorded for each value of the control parameters. The imaged area is $70 \times 70$~mm$^2$, with the left border close to the wavemaker edge. The surface-pattern distance is chosen as $h_p = 7$~mm in order to avoid light ray crossings due to strong surface curvature. The displacement field is determined using interrogation windows of size $16 \times 16$~pixels with an overlap of 8~pixels, resulting in a spatial resolution of approximately $0.5$~mm. In this situation, the surface height can be reconstructed within a precision of 2\%, yielding a vertical resolution of order of 2~$\mu$m.

\section{Longitudinal waves}
\label{sec:Longitudinal}

\begin{figure}
   \centerline{\includegraphics[width=16cm]{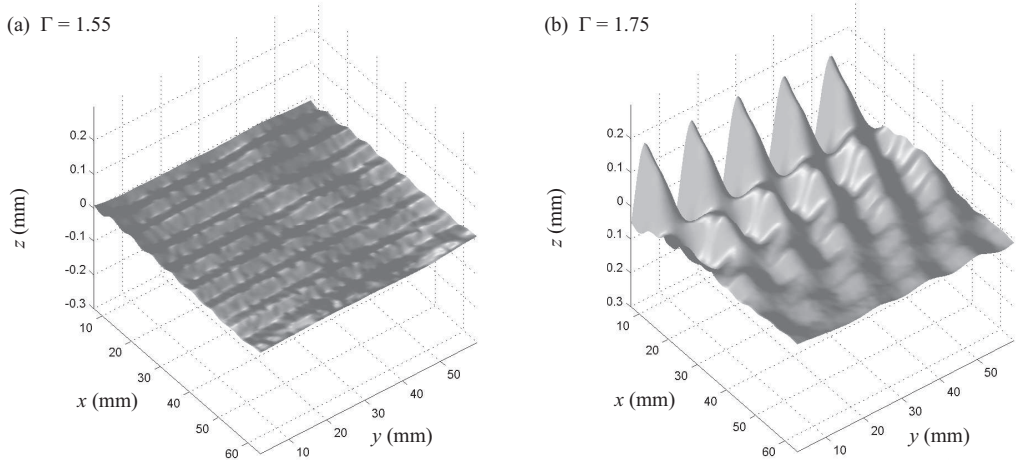}} 
  \caption{Three-dimensional views of the reconstructed surface height, as measured by FS-SS. The wavemaker of length $L=240$~mm is located at $x=0$ (upper left), with an oscillation frequency $f=40$~Hz and immersion depth $h_i=4$~mm. (a) Longitudinal waves for dimensionless acceleration $\Gamma = a_0\omega^2/g=1.55$ (4\% below the cross-wave threshold); (b) cross-waves for $\Gamma=1.75$ (9\% above the threshold). \label{fig:SurfaceShape}}
\end{figure}

Two examples of the reconstructed surface height are shown in figure~\ref{fig:SurfaceShape}. For low forcing amplitude $a_0$, only weak longitudinal waves are observed (figure \ref{fig:SurfaceShape}a), propagating in the $x>0$ direction. These longitudinal waves show an almost constant amplitude of the order of few $\mu$m along the wave crest ($y$ direction) and an attenuation along $x$.  As the forcing amplitude is increased, cross-waves are also generated (figure \ref{fig:SurfaceShape}b), oscillating at a frequency half of the wavemaker frequency, with crests nearly perpendicular to the plate. The cross-wave amplitude is much larger than that of the longitudinal wave, and the latter are visually difficult to detect once the cross-waves are present. We first focus in this section on the longitudinal waves, whereas the cross-waves are described in detail in the next section.

\begin{figure}
\centerline{\includegraphics[width=7.3 cm]{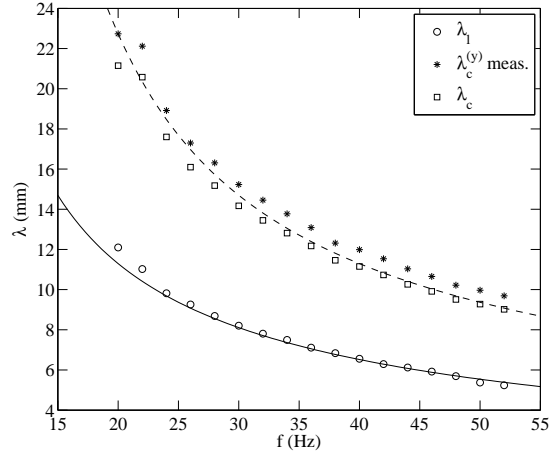}}
\caption{Wavelength of the primary longitudinal wave ($\circ$) and cross-wave ($\star$, $\square$) as a function of the forcing frequency.  Continuous line: prediction from the dispersion relation (\ref{eq:Dispersion}), from which the surface tension $\gamma \simeq 60 \times 10^{-3}$ N~m$^{-1}$ is determined. Dashed line: prediction from the dispersion relation for the frequency $f/2$, for the same value of $\gamma$. See Sec.~\ref{sec:sts} for the cross-wave wavelengths $\lambda_c$ and $\lambda_c^{(y)}$.  \label{fig:lambda}}
\end{figure}

The wavelengths $\lambda_\ell$ of the primary longitudinal waves, as measured from the FS-SS method, are shown in figure~\ref{fig:lambda} as a function of the forcing frequency. They are compared with the wavelength $\lambda = 2\pi / k$ deduced by inverting the dispersion relation for deep water waves,
\begin{equation}
\label{eq:Dispersion}
\omega^2 = g k + \frac{\gamma k^3}{\rho}.
\end{equation}
An excellent agreement is obtained if the water surface tension is taken equal to $\gamma = (60 \pm 2) \times 10^{-3}$ N~m$^{-1}$, a reasonable value for tap water.  This provides an accurate {\it in situ} determination of the surface tension, which is useful in the following for comparison with the wavelength of the cross-waves.

\begin{figure}
\centerline{
\includegraphics[width=8cm]{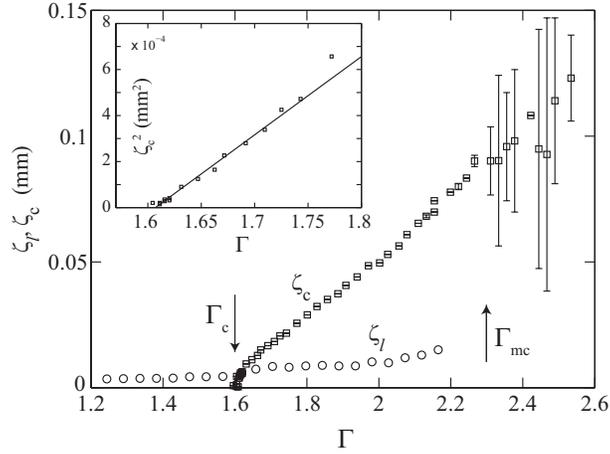}} 
\caption{Amplitude of the longitudinal wave $\zeta_\ell$ ($\circ$) and cross-wave  $\zeta_c$ ($\square$), measured at a fixed distance $x_0 = 20$~mm from the wavemaker, as a function of the dimensionless wavemaker acceleration $\Gamma = a_0\omega^2/g$, for $L=120$~mm, $f= 40$ Hz and $h_i = 4$ mm. The error bars reflect the slow modulation of the cross-wave pattern for a forcing acceleration above the secondary threshold $\Gamma_{mc}$. Inset: squared amplitude in the vicinity of the cross-wave transition, showing the classical supercritical behaviour $\zeta_c^2 \propto \Gamma-\Gamma_c$.}
\label{fig:AmplitudeVerticale}
\end{figure}

The surface profile can be described in general in terms of near-field and far-field contributions.\cite{Havelock1929,Mahony1972,Jones1984} It may be approximately written as a sum of 3 terms,
\begin{equation}
\label{eq:AmpLongitudinale}
\eta(x,y,t) = \zeta_0(x) \cos (\omega t) + \zeta_\ell(x) \cos (k_\ell x-\omega t) + \zeta_c(x,y,t).
\end{equation}
$\zeta_0(x)$ is a non-propagating local oscillation  at frequency $f$, induced by the flow
between the immersed wavemaker and the free surface; $\zeta_\ell(x)$ is the propagating longitudinal wave, also at frequency $f$; finally, 
$\zeta_c(x,y,t)$ is the cross-wave component at frequency $f/2$, which is present only for sufficiently large forcing acceleration (further described in the next section).

Sufficiently far from the wavemaker, the amplitude of the local oscillation $\zeta_0(x)$ may be neglected, so that the envelope of the longitudinal wave can be extracted as $\zeta_\ell(x) = \sqrt{2} \langle \langle \eta \rangle_y^2 \rangle_t^{1/2}$, where $\langle \cdot \rangle_y$ and $\langle \cdot \rangle_t$ denote the $y$ and time averages respectively. Averaging first over $y$ allows for measurement of $\zeta_\ell$ even when cross-waves are present in the field, since their oscillations in the $y$ direction cancel out.

The amplitude $\zeta_\ell$, measured at a fixed distance $x_0=20$~mm from the wavemaker, is plotted in figure \ref{fig:AmplitudeVerticale} as a function of the normalized acceleration $\Gamma = a_0 \omega^2/g$, at a fixed forcing frequency $f = 40$~Hz. Very low amplitudes of $\zeta_\ell$ are obtained, increasing from 4 to 15~$\mu$m as the forcing is increased.
The damping of the longitudinal wave in the $x$ direction can be estimated by fitting the wave envelope  by an exponential decay, $\zeta_\ell(x) \propto \exp(-x/\Lambda_\ell)$. The fitted attenuation length $\Lambda_\ell$ is found to decrease from 120 to 30~mm as the frequency is increased between 20 and 60~Hz. These values are approximately a factor of 2 below the expected attenuation length for linear waves in the limit of low viscosity,\cite{Leblond1987}
\begin{equation}
\Lambda_\ell \approx {\pi f}/ (\nu k^3).
\label{eq:leblond}
\end{equation}
The stronger dissipation observed in the experiment may originate from slight impurities on the surface, which are known to have a large influence on the damping of capillary waves.\cite{Stenvot1988}

\section{Cross-waves}
\label{sec:Cross-waves}

\subsection{Cross-wave bifurcation and growth time}

\begin{figure}
\centerline{\includegraphics[width=8cm]{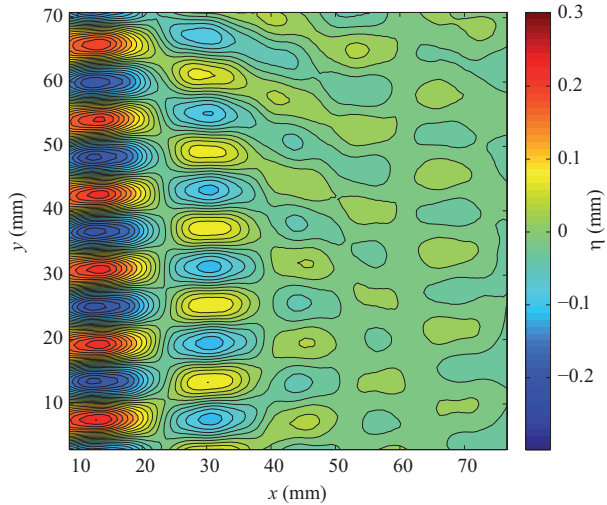}}
\caption{(Color online) Snapshot of the cross-wave pattern at an arbitrary time, obtained for a wavemaker immersion depth $h_i = 4$~mm, length $L=240$~mm, frequency $f = 40$~Hz, and $\Gamma / \Gamma_c = 1.17$. The wavemaker is along $y$ at $x = 0$ (slightly outside the imaged area), and centered on $y = 0$. The iso-levels lines correspond to  $0.02$~mm variations in the surface elevation.
\label{fig:isolevel}}
\end{figure}

We now turn to the cross-waves, which appear at sufficiently large forcing amplitude. Cross-waves oscillate with a frequency half the forcing frequency, as expected for parametrically excited waves. A snapshot illustrating the typical staggered structure of the cross-wave field is shown in figures~\ref{fig:SurfaceShape}(b) and \ref{fig:isolevel}, characterized by an oscillation of nearly constant amplitude along $y$, and a strongly damped oscillation along $x$.
Interestingly, the $y$-oscillation is stationary near the center of the wavemaker,  whereas a propagation is observed along the $x$-oscillation. Visual inspection shows that the two cross-wave systems emitted on each side of the wavemaker are in phase, i.e. with coinciding maxima and minima for $x<0$ and $x>0$.  This observation contrasts with the findings of Taneda,\cite{Taneda1994} in which the wavemaker were crossing the free surface and the two cross-wave systems on each side of the wavemaker were found to be out of phase (see also the sketch in the early paper of Faraday \cite{Faraday1831}). The wavemaker being totally immersed provides here a natural synchronization of the cross-waves on each side.

We first focus on the amplitude of the cross-waves as a function of the forcing amplitude, measured here from the laser beam deflection method at a distance $x_0 = 20$~mm from the wavemaker. For a frequency $f=40$~Hz and immersion depth $h_i = 4$~mm, the cross-wave amplitude  shows a classical bifurcation behavior, with a well defined threshold $\Gamma_c \simeq 1.61 \pm 0.02$ (Fig.~\ref{fig:AmplitudeVerticale}).  No hysteresis is found at the transition as the forcing acceleration is increased or decreased. For $\Gamma > \Gamma_c$, the amplitude compares well with the supercritical bifurcation law, $\zeta_c^2 \propto \Gamma - \Gamma_c$ (see the inset of Fig.~\ref{fig:AmplitudeVerticale}).  Slightly above the threshold, the amplitude of cross-waves reaches 0.1~mm, which is typically 5 to 10 larger than the amplitude of the longitudinal waves at the same distance $x_0$ from the wavemaker. Closer to the wavemaker, the cross-wave-to-longitudinal amplitude ratio is even larger, reaching values up to 50.

The pure cross-wave pattern of stationary amplitude is only found in a restricted acceleration range above the onset, for $\Gamma < \Gamma_{mc} \simeq 2.30 \pm  0.05$.
Above this secondary threshold, the cross-wave amplitude becomes slowly modulated in time, with a modulation frequency $f_m$ increasing from 0.2 to about 1.5 Hz as $\Gamma$ is increased above $\Gamma_{mc}$. This slow modulation is represented by the error bars in Fig.~\ref{fig:AmplitudeVerticale}, which are determined from the maximum and minimum of the oscillation amplitude during the modulation. The ratio $\Gamma_{mc} / \Gamma_{c}$ is found to depend both on the forcing frequency and the wavemaker length (one has $\Gamma_{mc} / \Gamma_{c} \simeq 1.5$ for $L=120$~mm, and 1.2 for $L=240$~mm). This slowly modulated cross-wave pattern at large forcing acceleration is consistent with other experiments performed with partially immersed wavemakers.\cite{Taneda1986, Lichter1986, Lichter1987, Shemer1989}  For even larger amplitude, the wave structure becomes disordered, and spatiotemporal defects gradually appear. The description of these time-dependent cross-wave patterns are beyond the scope of the present paper, and we restrict our attention in the following to the cross-wave of stationary amplitude.

\begin{figure}
\centerline{
\includegraphics[width=8cm]{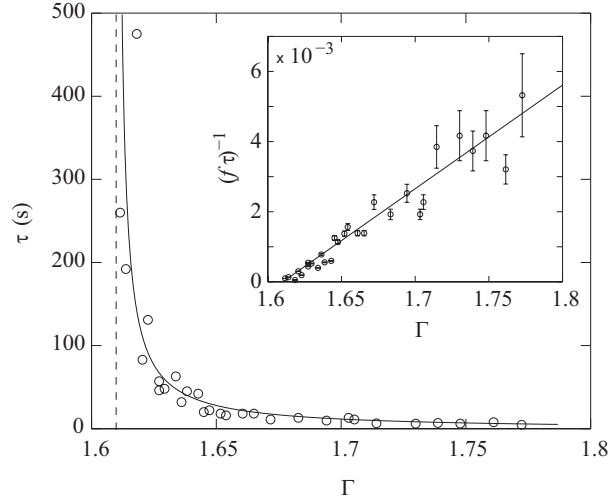}}
\caption{Growth time of the cross-waves, determined from the reflected laser beam method, as a function of the acceleration $\Gamma$. The line shows a best fit with $\tau \propto (\Gamma-\Gamma_c)^{-1}$, with $\Gamma_c = 1.61$. Inset: Dimensionless growth  rate $\sigma=(f\tau)^{-1}$, with the linear fit $\Gamma-\Gamma_c$. Error bars indicate a measurement uncertainty of 1~s. Wavemaker of length $L = 120$~mm,  $f=40$~Hz,  immersion depth $h_i = 4$ mm. 
\label{fig:Growthrate}}
\end{figure}

A remarkable feature of the cross-waves is their very long growth time, which may exceed 5 minutes (i.e. more than 20~000 forcing periods) close to the onset. This is in strong contrast with the longitudinal waves, which appear almost instantaneously as the wavemaker oscillation is started.
Interestingly, similar slow growths are also encountered in classical Faraday experiments,\cite{Benjamin1954,Shemer1989,Barnard1972} so this property seems to be generic for parametrically excited surface waves.

Estimates for the growth time  have been determined using the reflected laser beam method as follows. The forcing acceleration is increased, from a value slightly below the threshold, $0.99 \Gamma_c$, up to a final value between $\Gamma_c$ and $1.1 \Gamma_c$. The growth time $\tau$ is arbitrarily defined as the time at which the cross-wave reaches a significant amplitude of a few $\mu$m. In spite of this crude detection method, well defined growth rates are found, with a clear divergence near the threshold, as shown in Fig.~\ref{fig:Growthrate}. The non-dimensional growth time, $\sigma=1/(f\tau)$, shows a linear behavior close to the transition (inset of Fig.~\ref{fig:Growthrate}), here again a classical result for a supercritical bifurcation.

\begin{figure}
\centerline{\includegraphics[width=8cm]{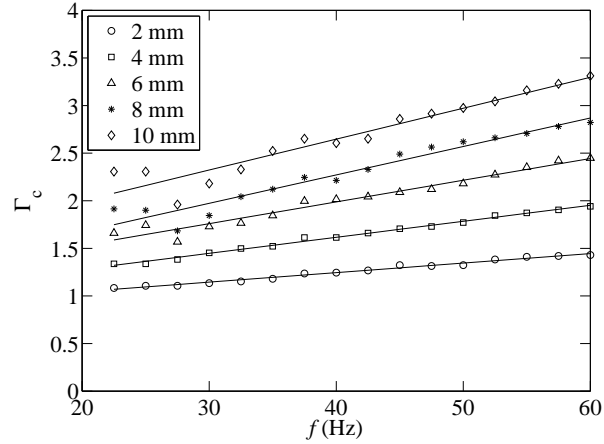}}
\caption{Critical acceleration $\Gamma_c$ for the onset of cross-waves as a function of the forcing frequency $f$ for various immersion depths $h_i$ between 2 and 10~mm. Lines are best linear fits. The scatter in $\Gamma_c$ at low frequency originates from residual resonances of the mechanical frame.
\label{fig:Onset}}
\end{figure}

For the explored ranges of frequency and immersion depth, critical wavemaker accelerations $\Gamma_c$ are of the order $1-4$ (Fig.~\ref{fig:Onset}).   As expected, as the immersion depth $h_i$ is increased, the influence of the wavemaker weakens, resulting in an increase in $\Gamma_c$ (this increase follows approximately the law $\Gamma_c \propto h_i^{0.5 \pm 0.1}$). Similarly, we have also observed that using a thinner wavemaker shifts $\Gamma_c$ towards larger values.  It is worth noting that these critical accelerations are much larger, at least by a factor of 10, than the prediction $\Gamma_{c,F} =4\nu \omega k/g$ for homogeneous Faraday waves in a fluid of low viscosity.\cite{Landau1976, Bechhoefer1995} Such large values of $\Gamma_c$ illustrate the low efficiency of the immersed wavemaker forcing compared to the homogeneous forcing through a modulated gravity in the classical Faraday configuration. The increase of $\Gamma_c$ with the frequency, also present for Faraday waves (one has $\Gamma_{c,F} \propto \omega^{5/3}$ at large frequency), originates from the nature of the forcing: since the wavemaker induces an acceleration of the free surface, it is naturally more efficient for gravity waves than for capillary waves.

\subsection{Spatiotemporal structure of the cross-waves}
\label{sec:sts}

\begin{figure}
\centerline{\includegraphics[width=10cm]{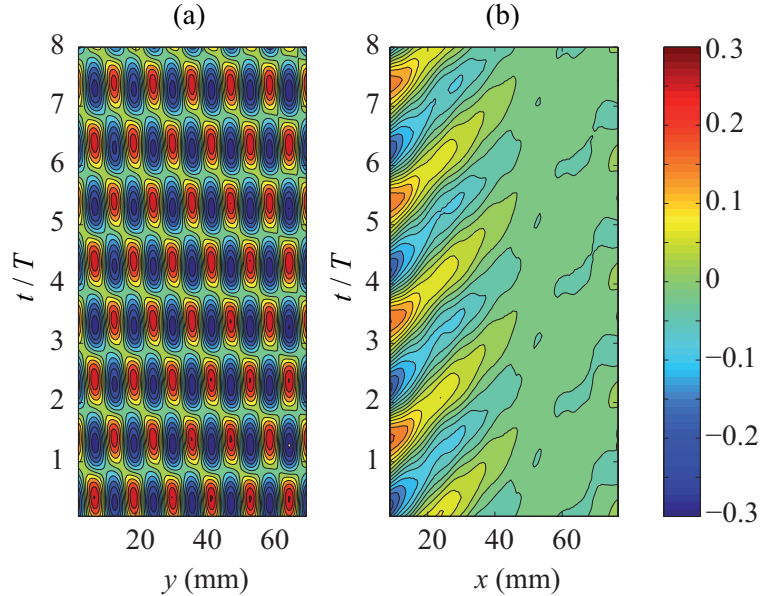}}
\caption{(Color online) (a) Spatiotemporal diagram $(y,t)$ at fixed $x_0 = 8$~mm, showing the stationarity  of the cross-wave along the wavemaker edge. (b) Spatiotemporal diagram $(x,t)$ at $y_0=0$, showing the propagation of the cross-waves in the $x>0$ direction. Time is normalized by the wavemaker period $T = 1/f$, showing the period $2T$ of the subharmonic wave. Same parameters as in Fig.~\ref{fig:isolevel}.
\label{fig:mxy}}
\end{figure}

\begin{figure}
\centerline{\includegraphics[width = 4cm]{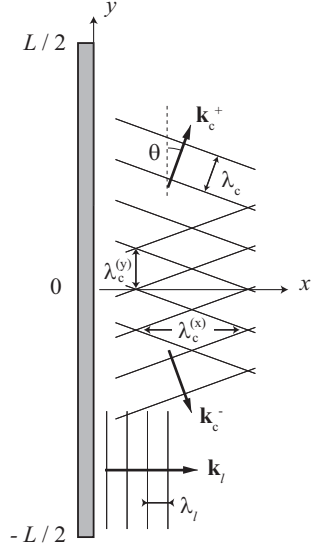}}
\caption{Sketch of the two propagative parametric components ${\bf k}_c^+$ and ${\bf k}_c^-$ of the cross-wave, making an angle $\theta$ with the wavemaker. The interference of these two waves forms a staggered pattern, stationary along $y$ and propagative along $x$. Also shown is the primary wave ${\bf k}_{l}$ propagating along $x$.
\label{fig:sketch_pattern}}
\end{figure}

We now describe in more detail the spatiotemporal structure of the cross-wave pattern. When observed at a constant distance $x$ from the oscillating plate, the surface height profile along $y$ shows a stationary wave oscillating at half the forcing frequency, as shown in the $(y,t)$ spatiotemporal diagram in Fig.~\ref{fig:mxy}(a).  However, when observed at a constant $y$ position, the pattern now corresponds to a strongly damped propagating wave moving away in the $x>0$ direction, as shown in the $(x,t)$ spatiotemporal diagram in Fig.~\ref{fig:mxy}(b). This suggests a local description of the cross-wave pattern in terms of the superposition of two oblique propagative waves of frequency $f/2$, described by the two symmetric wavevectors $\mathbf{k}_c^+$ and $\mathbf{k}_c^-$ making an angle $\theta$ with the wavemaker (see Fig.~\ref{fig:sketch_pattern}).
Accordingly, the cross-wave field may be written, in the limit of a long wavemaker,
\begin{equation}
\label{eq:AmpCross}
\zeta_c(x,y,t) = \zeta_c(x) \left[\cos (\mathbf{k}_c^+ \cdot {\bf r} -\frac{\omega}{2}t) + \cos (\mathbf{k}_c^- \cdot {\bf r} -\frac{\omega}{2}t)\right],
\end{equation}
with $\mathbf{k}_c^\pm = {k_c} (\sin \theta \mathbf{e}_x \pm \cos \theta \mathbf{e}_y)$, ${\bf r} = x \mathbf{e}_x  + y \mathbf{e}_y$, and $\zeta_c(x)$ the cross-wave envelope. A pure stationary cross-wave, invariant along $x$, would correspond to $\mathbf{k}_c^\pm = \pm k_c {\bf e}_y$, and hence $\theta=0$.\cite{Mahony1972,Jones1984}

In order to determine the wavevectors $\mathbf{k}_c^\pm$ from the surface height, we have measured the apparent wavelengths, $\lambda_c^{(x)} = 2 \pi / k_{cx}$ and $\lambda_c^{(y)} = 2 \pi / k_{cy}$, simply defined as the distance between wave extrema along each direction. The apparent wavelength $\lambda_c^{(y)}$ can be precisely determined to within 1\%, whereas the measurement of $\lambda_c^{(x)}$ suffers from a much larger uncertainty due to the strong damping along $x$, and could be determined only to within 10\%. From those measurements, the true wavelength of
the oblique propagating components is defined as $\lambda_c = 2 \pi / k_c$, with $k_c = (k_{cx}^2 + k_{cy}^2)^{1/2}$.

Figure~\ref{fig:lambda} shows, in addition to the wavelength of the longitudinal waves discussed in Sec.~\ref{sec:Longitudinal}, the measured apparent wavelength $\lambda_c^{(y)}$ and true wavelength $\lambda_c$ of the cross-waves as a function of the forcing frequency. For each value of the frequency, the acceleration $\Gamma$ has been set 10\% above the critical amplitude $\Gamma_c(f)$, so that the measurements are obtained following closely the stability curve of Fig.~\ref{fig:Onset}. The cross-wave wavelength $\lambda_c$ is found in excellent agreement with the prediction of the dispersion relation (\ref{eq:Dispersion}) for a frequency $f/2$ (dashed line),  confirming that the cross-wave pattern can be described in terms of the interference of two oblique parametric propagating waves.
We note that the number of cross-wave crests and troughs along the wavemaker lies here in the range 10 to 24, so discretization effects of order of 5 to 10\% could be expected. However, the smooth variation of $\lambda_c$ with the frequency indicates that no noticeable discretization effect could be detected, which is probably a consequence of the absence of rigid boundary conditions at the water surface at each end of the wavemaker.

Finally, the cross-wave envelope $\zeta_c(x)$ can be extracted from Eq.~(\ref{eq:AmpCross}), following a similar procedure as for the longitudinal wave. An exponential fit of the envelope, $\zeta_c(x) \propto \exp(-x / \Lambda_{cx})$, yields apparent attenuation length $\Lambda_{cx}$ decreasing from 35 to 18~mm  as the forcing frequency is increased. This attenuation length is of order of $\lambda_{cx}$, with an approximately constant ratio $\Lambda_{cx} / \lambda_{cx} \simeq 0.7 \pm 0.1$ in the explored range of frequencies, confirming the strongly damped nature of the cross-waves. This damping is actually too strong to be associated with viscous attenuation: if we compute the attenuation length according to Eq.~(\ref{eq:leblond}), and take into account the projection of the wavevectors ${\bf k}_c^\pm$ with respect to $x$, we would have an apparent attenuation length $\Lambda_c \sin \theta$ ranging between 1.2~m and 0.16~m for the range of frequency considered here, i.e. about 30 times larger than the measured $\Lambda_{cx}$. This large discrepancy clearly shows that cross-waves are intrinsically confined in the vicinity of the wavemaker, and that this confinement cannot be related to viscous damping. Cross-waves are therefore essentially excited from a resonance which involves the local structure of the flow close to the wavemaker: they are sustained in the region where the non-propagating oscillation $\zeta_0(x)$ [see Eq.~(\ref{eq:AmpLongitudinale})] is significant, which acts as waveguide along the wavemaker, and are evanescent outside this region.

\subsection{Finite length effects}

\begin{figure}
\centerline{\includegraphics[width=8cm]{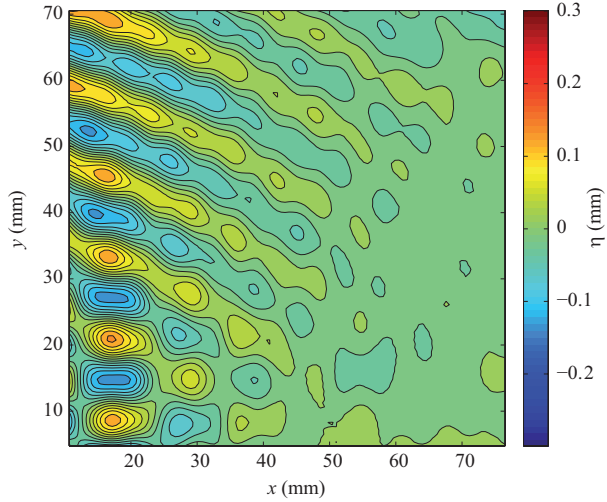}}
\caption{(Color online) Cross-waves showing finite size effects, for a short wavemaker of length $L=120$~mm.  The wavemaker is at $x = 0$, centered on $y = 0$, with its upper end at $y=L/2=60$~mm. The upward propagating parametric wave ${\bf k}_c^+$ is stronger than the downward wave ${\bf k}_c^-$ near the wavemaker end, resulting in a nearly symmetric stationary wave at $y \simeq 0$ but a pure propagative wave near $y = L/2$ and above.
Same parameters as in Fig.~\ref{fig:isolevel}.
\label{fig:isoshort}}
\end{figure}

The previous description of the cross-wave pattern as an interference between two plane propagative waves at frequency $f/2$ implicitly assumes that the wavemaker is infinite. For the long wavemaker of length $L=240$~mm, a nearly symmetric staggered pattern is actually observed in the imaged region near the center (see Fig.~\ref{fig:isolevel}). However, closer to the wavemaker ends, the cross-wave pattern is no longer symmetric, as illustrated by the wave field in figure~\ref{fig:isoshort}, obtained for a shorter wavemaker of length $L=120$~mm and a measurement window located at $0 \leq y \leq L/2$. Whereas a symmetric pattern is still observed near the center, at $y \simeq 0$, nearly pure propagative waves ${\bf k}_c^\pm$ with oblique crests are observed near the ends, and ``escape'' far from the wavemaker in the $y$ direction.

Figure~\ref{fig:isoshort} suggests that, now, the amplitudes of the two propagative components of the cross-wave are no longer equal near the ends of the wavemaker.  Accordingly, the cross-wave field (\ref{eq:AmpCross}) should be generalized as follows,
\begin{equation}
\label{eq:AmpCross_apam}
\zeta_c(x,y,t) = \zeta_c^+(x,y) \cos (\mathbf{k}_c^+ \cdot {\bf r} -\frac{\omega}{2}t) + \zeta_c^-(x,y) \cos (\mathbf{k}_c^- \cdot {\bf r} -\frac{\omega}{2}t),
\end{equation}
where the two envelopes $\zeta_{c}^\pm$ now depend on both $x$ and $y$. For $\zeta_c^+ \simeq \zeta_c^-$, the resulting pattern is symmetric, stationary along $y$ and propagating along $x$, whereas the envelopes should strongly differ near the wavemaker ends, leading to pure propagative waves.

\begin{figure}
\centerline{\includegraphics[width=7cm]{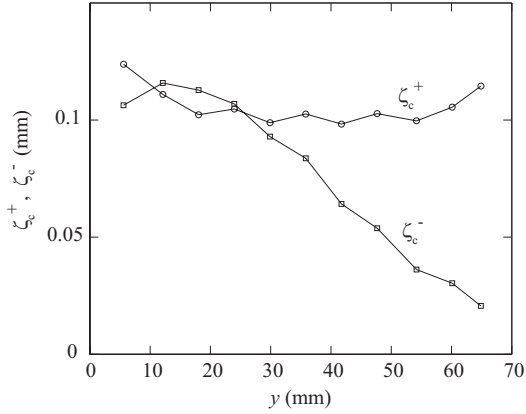}}
\caption{Amplitude of the two propagative components of the cross-wave, $\zeta_c^+$ and $\zeta_c^-$, in the upper half of the short wavemaker $0 \leq y \leq L/2$, with $L=120$~mm. The two amplitudes are nearly equal at the center of the wavemaker, at $y \simeq 0$, resulting in a stationary cross-wave pattern along $y$. Near the upper end, at $y \simeq L/2$, the downward amplitude $\zeta_c^-$ vanishes, resulting in an approximately pure upward propagating wave.} \label{fig:apam}
\end{figure}

The two envelopes $\zeta_{c}^\pm$ have been extracted from the time series of the cross-wave height profile along $y$ at a fixed distance $x_0$ from the wavemaker. Figure~\ref{fig:apam} shows that these two envelopes are indeed nearly equal near the wavemaker center, but the wave amplitude $\zeta_c^-$ is found to decrease and is almost zero at the wavemaker end $y \simeq L/2 =  60$~mm. The approximately linear decrease of $\zeta_{c}^-$ suggests that the downward wave amplitude at a given location $y$ is the result of the sum of elementary wave sources from $L/2$ to $y$, and similarly for the upward wave from $-L/2$ to $y$.  Accordingly, for a very long wavemaker, the two wave components have essentially constant amplitudes everywhere except near the two ends, resulting in a symmetric staggered pattern far from the ends.

\subsection{Selection of the cross-wave angle}

\begin{figure}
\centerline{
\includegraphics[width=14cm]{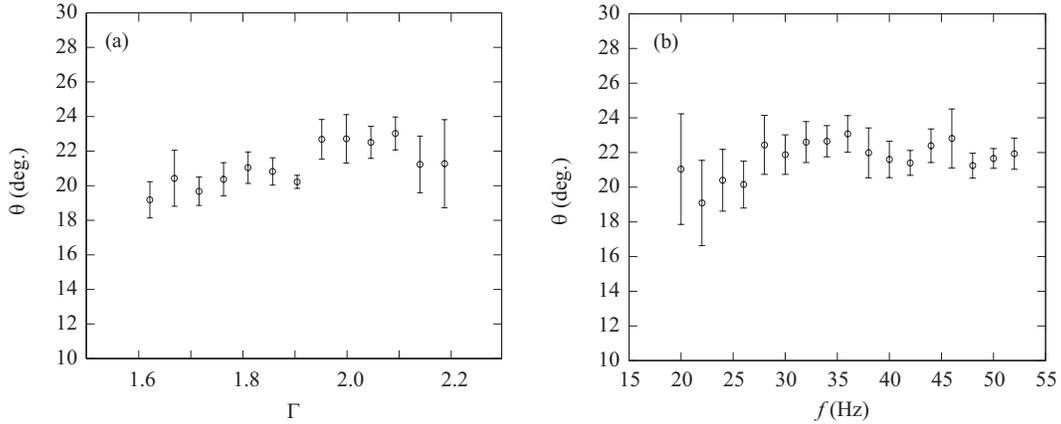}}
  \caption{(a) Cross-wave angle $\theta$ as a function of the forcing acceleration $\Gamma$, for a fixed forcing frequency $f = 40$ Hz. (b) Angle $\theta$ as a function of the forcing frequency $f$ for an acceleration 10\% above the corresponding cross-wave threshold $\Gamma_c(f)$.
  \label{fig:Angle}}
\end{figure}

We finally characterize the angle $\theta$ between the two wavenumbers ${\bf k}_{c \pm}$ and the wavemaker.
From the measurements of the two apparent wavelengths $\lambda_c^{(x)}$ and $\lambda_c^{(y)}$, this angle is simply defined as $\theta = \tan^{-1}(\lambda_c^{(y)} / \lambda_c^{(x)})$. For the wavemaker of length $L=240$~mm, with an immersion depth $h_i = 4$~mm, the angle $\theta$ is plotted in Fig.~\ref{fig:Angle}(a) as a function of $\Gamma$ for fixed $f=40$~Hz, and in Fig.~\ref{fig:Angle}(b) as a function of $f$ following the stability curve, for $\Gamma \simeq 1.1 \Gamma_c(f)$. Remarkably, the angle
is found nearly constant, in the range $19-23^\circ$, for the accelerations and frequencies considered here. Note that the angles obtained here are much lower than the ones reported by Taneda\cite{Taneda1994} for a horizontally half-immersed cylinder.

The angle $\theta$ of the cross-wave pattern should result from a nonlinear resonance mechanism implying the two oblique parametric waves ${\bf k}_c^\pm$.\cite{Mahony1972,Jones1984}
A triadic interaction provides the simplest frame to describe this resonance: two parametric waves ${\bf k}_{c}^\pm$ may interact with a third wave ${\bf K}_0$ if the spatial and temporal resonance conditions are satisfied,\cite{McGoldrick1965}
\begin{eqnarray*}
 {\bf k}_c^+ + {\bf k}_c^- &=& {\bf K}_0, \\
 \omega({\bf k}_c^+) + \omega({\bf k}_c^-) &=& \omega({\bf K}_0).
\end{eqnarray*}
Since we have $\omega({\bf k}_c^+) = \omega({\bf k}_c^-) = \omega/2$,
these relations imply that $|{\bf k}_c^+|$ = $|{\bf k}_c^-|$, so that the
wavevector ${\bf K}_0$ must be along $x$ (see Fig.~\ref{fig:sketch_pattern}) and have frequency given by the forcing frequency $\omega$.  The angle $\theta$ must therefore satisfy  $2 | {\bf k}_c^\pm | \sin \theta = |{\bf K}_0|$. The particular case $\theta=0$ hence corresponds to a pure stationary cross-wave pattern, invariant along $x$, whereas one has $\theta \neq 0$ for the staggered  pattern considered here. The question is now to determine the physically relevant  third wavevector ${\bf K}_0$ responsible for the growth of the two oblique parametric waves ${\bf k}_c^\pm$.

A first natural choice for ${\bf K}_0$ is the wavevector ${\bf k}_\ell$ of the longitudinal primary propagative wave selected by the dispersion relation (see Fig.~\ref{fig:sketch_pattern}). Accordingly, the selected angle would be given by
\begin{equation}
\theta = \sin^{-1} \left( \frac{k(\omega)}{2 k(\omega/2)} \right),
\label{eq:th}
\end{equation}
where $k(\omega)$ is the inverse of the dispersion relation (\ref{eq:Dispersion}).
However, this choice would yield angles much larger than the experimental angles, starting from $90^\mathrm{o}$ for $f = (9/2)^{1/4} f_c \simeq 20.6$~Hz (where $f_c \simeq 14.2$~Hz is the capillary-gravity crossover frequency), and saturating towards $\theta_\infty = \sin^{-1} (2^{-1/3}) \simeq 52.5^\mathrm{o}$ for pure capillary waves at large frequency. A resonance mechanism between the two parametric waves and the primary longitudinal wave is therefore clearly ruled out by the present observation. This is another indication that the cross-waves are directly excited by the oscillating surface deformation $\zeta_0(x)$ above the wavemaker [see Eq.~(\ref{eq:AmpLongitudinale})], and not by the longitudinal wave.

Assuming now that the cross-wave resonance is governed by the oscillating flow close to the wavemaker, then the natural choice for ${\bf K}_0$ must be based on the characteristic length associated with the surface deformation above the wavemaker, $\zeta_0(x)$. 
It may be described in terms of a characteristic deformation length, $L_0 = 2\pi / K_0$, which should depend on the geometrical features of the wavemaker, such as the wavemaker width $w$ and the immersion depth $h_i$ (see Fig.~\ref{fig:setup}). Note that the viscous length associated with the Stokes layers on the wavemaker sidewalls, $\delta \simeq \sqrt{\nu / \omega}$, is less than 0.1~mm for the forcing frequencies considered here, and is not expected to contribute significantly to this deformation length.

It is clear that the interaction between the localized oscillation $\zeta_0(x)$ and the parametric waves ${\bf k}_c^\pm$ cannot be considered strictly as a true three-wave resonance: ${\bf K}_0$ is not a plane wave of infinite extent, since the surface disturbance remains essentially localized near the wavemaker.\cite{Mahony1972} The triadic resonance formalism must be therefore considered here as a convenient qualitative description of the true resonance mechanism.  For a given wavemaker geometry, a broad range of ${\bf K}_{0}$ should be excited, corresponding to the spectrum of the surface deformation above the wavemaker, leading to a continuum of possible resonant interactions. However, close to the cross-wave threshold, only the most energetic wavenumber should contribute to the resonance, which justifies the use of a discrete three-wave resonance framework.

Applying Eq.~(\ref{eq:th}) for $k(\omega) = 2\pi / L_0$ with $L_0$ of order of $15-20$~mm actually yields angles $\theta$ in the range $15-30^\mathrm{o}$, which is close to the measured values. This order or magnitude for $L_0$ compares reasonably with the wavemaker width $w=5$~mm and the immersion depth $h_i = 4$~mm considered here. However, this crude description would predict a decreasing angle $\theta$ as the frequency is increased, a behavior which is not observed in Fig.~\ref{fig:Angle}.  Although the agreement is only qualitative at this point, we can conclude that the resonance  responsible for the selection of the cross-wave angle must involve a characteristic length associated with the oscillating flow between the wavemaker and the free surface, and not with the primary longitudinal wave.

\section{Conclusion}

In this paper we have investigated the parametric instability leading to the generation of cross-waves excited by a fully immersed wavemaker. The use of a fully immersed wavemaker offers a great simplification of the forcing mechanism, by avoiding the complex dynamics of an oscillating meniscus. As a result, the generation of longitudinal primary waves is very weak, so it is possible here to investigate the structure of a nearly pure cross-wave pattern.

Free-Surface Synthetic Schlieren, an optical method for the time-resolved measurement of the  surface topography, proved useful to characterize in detail the spatiotemporal properties of the cross-wave. For the one-dimensional wavemaker investigated here, apart from finite length effects, the cross-wave pattern is well described by the interference of two parametric propagative waves ${\bf k}_c^+$ and ${\bf k}_c^-$ making an angle $\theta$ with the wavemaker. The resulting cross-wave pattern is not strictly normal to the wavemaker, but rather shows a staggered structure, stationary along the wavemaker and propagating normal to it. This cross-wave pattern is strongly confined near the wavemaker, confirming that it originates from a resonance with the local oscillating flow induced by the wavemaker. We may speculate that this staggered pattern generalizes the usual cross-wave pattern observed for gravity waves in large scale experiments, which is strictly stationary and hence corresponds to the specific case $\theta=0$.

Our results suggest that the resonance responsible for the growth of the cross-wave may be simply described by a three-wave interaction mechanism, in which the oscillating flow above the wavemaker is modeled by a pseudo third wavevector. This mechanism provides a qualitative agreement with the measured cross-wave angle $\theta$ if the third wavevector is based on a deformation length of the order of the characteristic dimensions of the wavemaker. The wavemaker geometry is therefore expected to play an important role in the selection of the cross-wave angle, which cannot be considered as a universal feature of capillary cross-waves. Further analysis of the flow between the wavemaker and the free surface, for wavemakers of different size, shape and immersion depth, is needed to support this picture.

\acknowledgments
We acknowledge A. Aubertin, L. Auffray, R. Pidoux and N. Schneider for experimental help, and P.-P. Cortet, M. Rossi and L. Tuckerman for fruitful discussions.

\end{document}